\documentclass{desyproc}

\newcommand{\ttbar}     {\ensuremath{t\bar{t}}}
\newcommand{\ppbar}     {\ensuremath{p\bar{p}}}
\newcommand{\Afb}       {\ensuremath{A_{FB}}}
\newcommand{\Al}        {\ensuremath{A_\ell}}
\newcommand{\All}        {\ensuremath{A_{FB}^{\ell\ell}}}
\newcommand{\Ac}        {\ensuremath{A_C}}
\newcommand{\Acll}      {\ensuremath{A_C^{\ell\ell}}}
\newcommand{\ljets}     {\ensuremath{\ell}+jets} 

\begin{document}
\title{Experimental status of top charge asymmetry measurements}

\author{{\slshape Viatcheslav Sharyy}  
for the D0, CDF, Atlas and CMS collaborations\\[1ex]
CEA, IRFU, SPP, Bat. 141, 91191 Gif-sur-Yvette, France}

\contribID{xy}  
\confID{7095}
\desyproc{DESY-PROC-2013-XY}
\acronym{TOP2013}
\doi            

\maketitle

\begin{abstract}
The latest measurements of the asymmetry in the angular distributions of the \ttbar\ events are reviewed.
The measurements of the forward-backward asymmetry \Afb\ in the \ppbar\ 1.98 TeV collision at the Tevatron 
show some tension with the standard model calculation, while  
results of the measurements of the charge asymmetry \Ac\ in $pp$ collisions 
at 7 TeV and 8 TeV at the LHC are compatible with standard model prediction. 
\end{abstract}

\section {Introduction}

The measurement of the asymmetry in the angular distributions of \ttbar\ events 
is a powerful test of the standard model (SM) predictions, and allows to probe for Physics beyond the SM.
Different asymmetries are considered at the Tevatron and LHC.
At the Tevatron, the \ttbar\ pairs are produced in the \ppbar\ collisions, so we can define 
the forward-backward asymmetry  as
\[ A_{FB} = \frac{N(\Delta y >0 ) - N(\Delta y <0 )}{N(\Delta y >0 ) + N(\Delta y <0 )}, \]
where $\Delta y = y_t - y_{\bar{t}}$ is a difference in rapidity of top and antitop quarks, 
$N(...)$ is a corresponding number of \ttbar\ events.
At the LHC, the $pp$ collisions are forward-backward symmetric, so the charge asymmetry
is considered:
\[ A_{C} = \frac {N(\Delta |y| >0 ) - N(\Delta |y| <0 ) }{ N(\Delta |y| >0 ) + N(\Delta |y| <0) }, \]
where $\Delta|y| = |y_t| -  |y_{\bar{t}}|$ is a difference in absolute
rapidity of top and antitop  quarks.  The latest SM calculations for those
quantities yield $ (8.7^{+0.6}_{-0.5})  \%$ for the Tevatron, $(1.23\pm
0.05) \%$ for the LHC 7 TeV and  $(1.11\pm 0.05) \%$ for the LHC 8 TeV
\cite{Bernreuther:2012sx}.    See   \cite{Westhoff:2013ixa}  for   the
discussion  about  details  of theoretical  predictions  and  possible
contributions from the non-SM processes.

The analysis of experimental data  includes several steps. After event
selection the top  and antitop quarks kinematic parameters  need to be
reconstructed using  measured parameters of leptons,  jets and missing
transverse energy.   Two final states  are usually considered  for the
asymmetry measurement: \ljets\ and  dileptons.  In \ljets\ final state
the \ttbar\  pair is  decaying to $t\bar{t}\to  W^+ W^-\  b\bar{b} \to
\ell\nu\  q\bar{q^\prime}\ b\bar{b}$,  where direction  and transverse
momentum ($p_T$) for lepton $\ell$  and four quark jets are sufficient
for reconstruction of the top and antitop parameters. Usually, more or
less  sophisticated  kinematic  reconstruction  methods  are  used  to
account  for  the ambiguity  in  attributing  jets to  partons.  These
methods usually  use constrains  on two-jets and  three-jets invariant
masses which correspond to the W-boson and to the top quark masses and
improve  the  uncertainty  on  the measured  jets  energies.   In  the
dilepton final state, the \ttbar\ pair  decays to the final state with
two  non-detected  neutrinos,  $t\bar{t}\to   W^+  W^-\  b\bar{b}  \to
\ell^+\ell^-\  \nu\bar{\nu}\ b\bar{b}$.   The  reconstruction in  this
final state requires a ``scan'' of  the phase space constrained by the
experimental measured  parameters of leptons and  jets.  Parameters of
the  top quark  and  antiquark are  calculated as  a  weighted sum  of
reconstructed parameters  in all  scan points.  The  last step  in the
asymmetry   measurements  is   a   unfolding   of  the   reconstructed
distributions to the parton  level.  Strictly speaking, such unfolding
is  not required  if  we  restrict the  measurement  to the  inclusive
asymmetry  only.  Usually,  it is  not the  case, because  it is  also
interesting  to measure  the asymmetry  dependence from  the invariant
mass of \ttbar\ pair ($m_{t\bar{t}}$), $\Delta y$ or other parameters.
Such differential  measurement is more  sensitive to the  possible new
physics contribution, since it is expected to contribute more in some
region or phase space, e.g. at high $m_{t\bar{t}}$.

The procedures of reconstruction and unfolding of top quark parameters
complicate quite a lot the asymmetry analyses and require a careful
calibration.  The alternative approach has been developed for the
asymmetry measurements. Instead of measuring the quark asymmetry, we
can measure asymmetry in the distributions of leptons. Since direction
of leptons is measured with a good precision, no top quark
reconstruction or unfolding is needed. The drawback of this approach,
that the leptonic asymmetry isn't as powerful as the top quark
asymmetry, because the direction of leptons is not fully correlated
with the direction of top quark.  For example, at the Tevatron, the
leptonic asymmetry is defined as
\[ 
A_\ell = \frac{N(q\cdot y_\ell > 0) - N(q\cdot y_\ell < 0)}{N(q\cdot
  y_\ell > 0) + N(q\cdot y_\ell < 0)},
\]
where $y_\ell$ and $q$ is a lepton rapidity and charge.  \Al\ is
predicted to be $(3.8\pm 0.6 \%)$ \cite{Bernreuther:2012sx}.  More
interestingly, it was found, that the measurement of the angular
distribution of leptons is complementary to the \ttbar\ asymmetry
measurement.  This is related to the fact, that the angular
distribution of leptons is affected not only by the angular
distribution of top quark but also by its polarization.  In the SM the
top quark polarization is zero, but could be significantly different
for the non-SM contribution, e.g.  in the \ttbar\ production via
axigluon mechanism \cite{Falkowski:2012cu}.  In the dilepton final
state, we also can measure the two-lepton asymmetry, constructed
analogously to the \ttbar\ asymmetry. It is defined at the Tevatron
as:
\[
A_{FB}^{ll} = \frac{N(\Delta y_\ell > 0) - N(\Delta y_\ell <
  0)}{N(\Delta y_\ell > 0) + N(\Delta y_\ell < 0)},
\]
and at the LHC:
\[ 
 A_{C}^{ll} = \frac{N(\Delta |y_\ell| > 0) - N(\Delta |y_\ell| <
   0)}{N(\Delta |y_\ell| > 0) + N(\Delta |y_\ell| < 0)},
\]
where difference in leptons rapidities is $\Delta y_\ell = y_{\ell^+}
- y_{\ell^-}$ and $\Delta |y_\ell| = |y_{\ell^+}| - |y_{\ell^-}|$.

\section{LHC results}
\begin{figure}
\begin{minipage}[t]{0.49\textwidth}
\includegraphics[width=\textwidth]{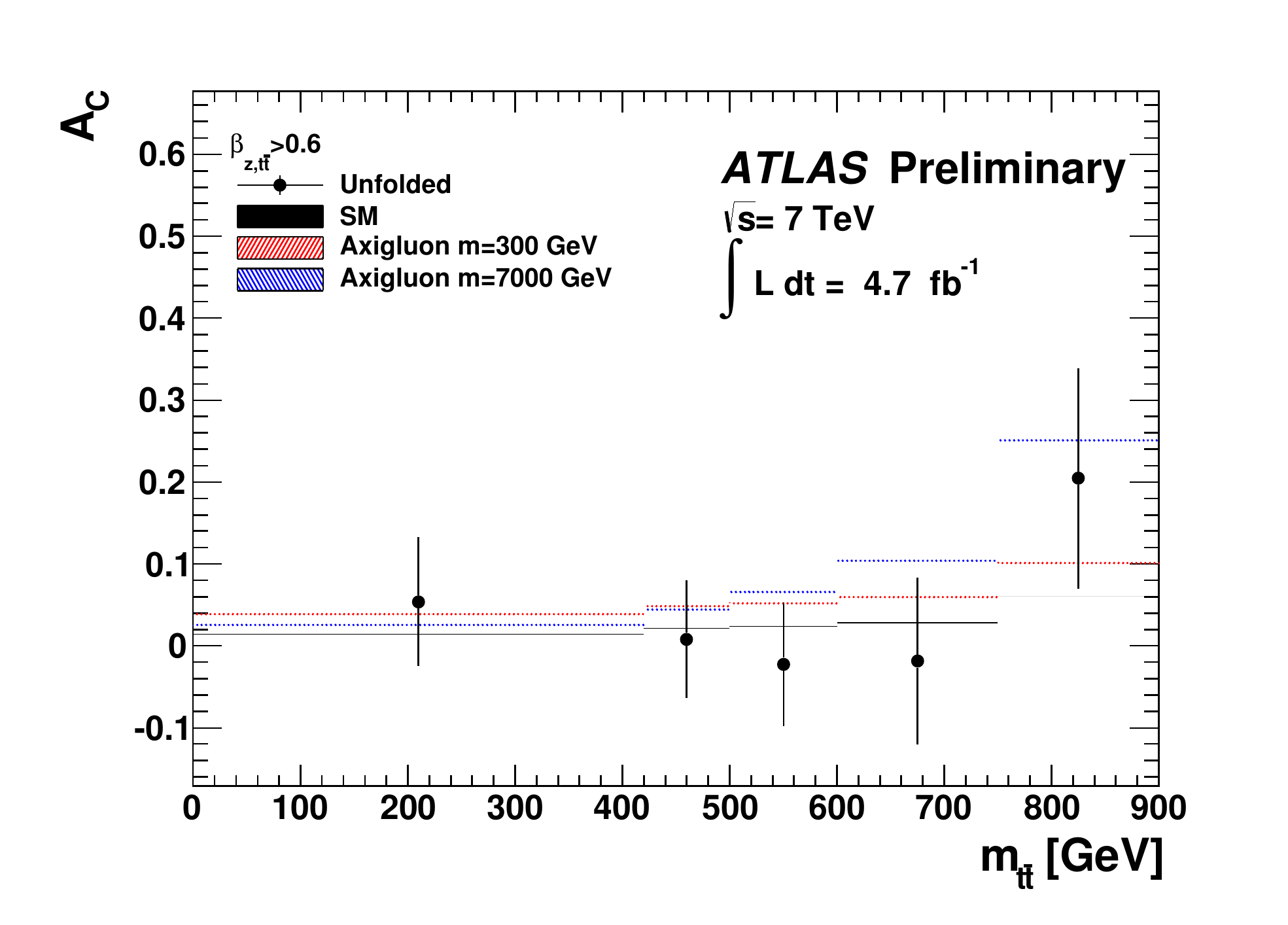}
\caption{\label{Fig:atlas1}Charge asymmetry distribution as a function
  of $m_{t\bar{t}}$ for the events with \ttbar\ velocity~$>0.6$ as
  measured by the Atlas experiment~\cite{ATLAS:ljets}.}
\end{minipage}
\hfill
\begin{minipage}[t]{0.46\textwidth}
\centerline{\includegraphics[width=\textwidth]{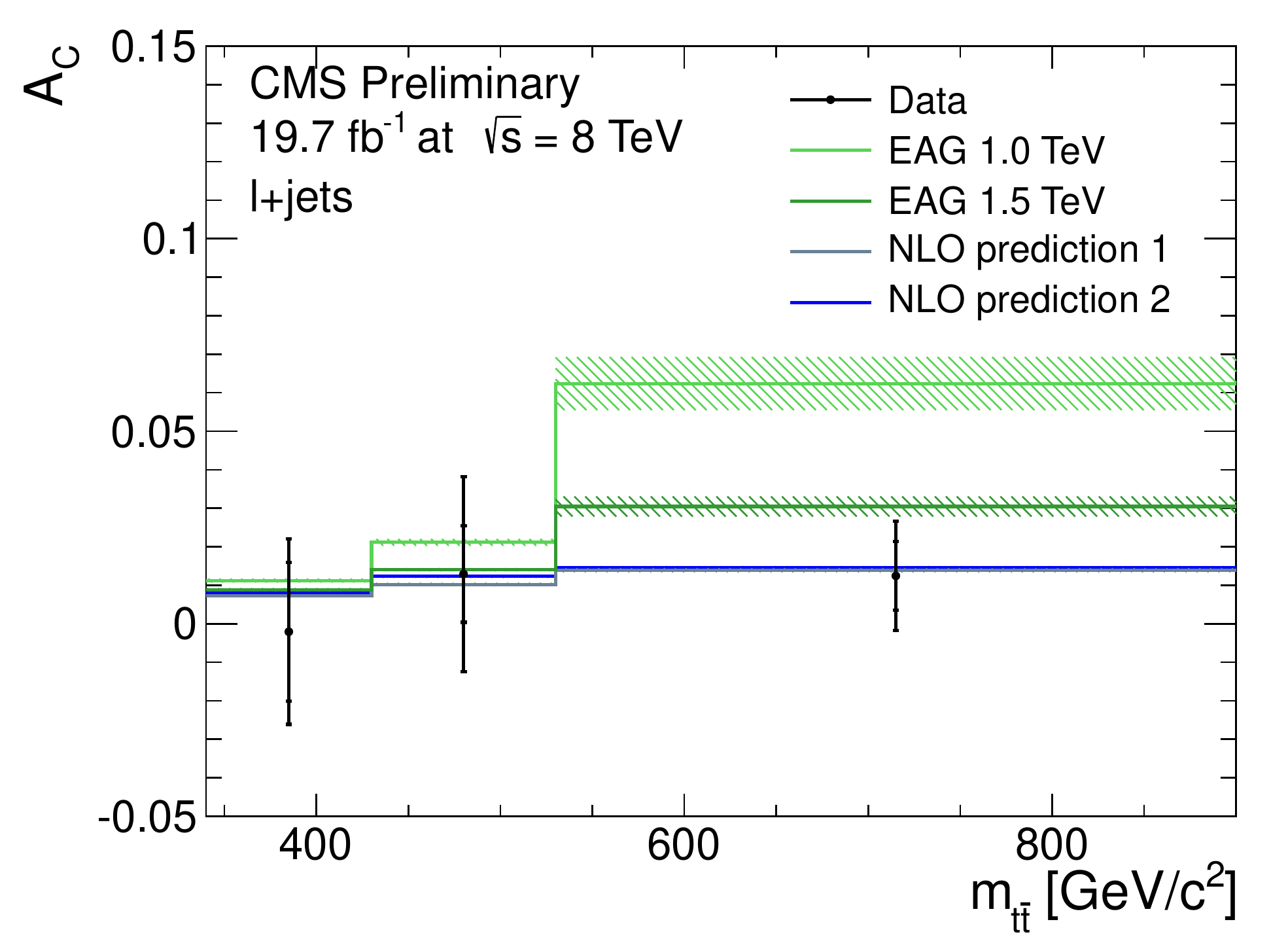}}
\caption{\label{Fig:cms1}Charge asymmetry distribution as a function
  of $m_{t\bar{t}}$ as measured by the CMS
  experiment~\cite{CMS:ljets}.}
\end{minipage}
\end{figure}

The results of the \Ac\ measurement by ATLAS at 7
TeV~\cite{ATLAS:ljets,ATLAS:dileptons} as well as measurements by CMS
at 7~TeV~\cite{CMS:ljets, CMS:dileptons} and 8~TeV~\cite{CMS:8tev} are
shown in the Table~\ref{LHC_results}.  The measured values are all
compatible with each other and with SM predictions. Measurements of
the leptonic asymmetry \Acll,~Table~\ref{LHC_results_lepton}, also
don't show any deviation from the SM expectation.
The enormous statistics accumulated at the LHC allows to investigate
the restricted phase space regions, e.g. high velocity and high
$m_{\ttbar}$ mass regions.  In both cases the expectation for the SM
asymmetry is larger than for the inclusive asymmetry and the possible
contribution from the non SM physics are also expected to be enhanced,
see e.g.~\cite{AguilarSaavedra:2012va}.  Both experiments have looked
at the asymmetry differential distributions, but no deviation from SM
has been found.  For illustration, see two selected distributions in
Fig.~\ref{Fig:atlas1},\ref{Fig:cms1}.

\begin{table}
\renewcommand{\arraystretch}{1.3}
\begin{tabular}{l|c|c|c|c|c}
\hline \multicolumn{4}{c|}{Measurement} & Measured Value, \% &
\parbox{2.7cm}{Theoretical \\ Expectation\cite{Bernreuther:2012sx},\%}
\\ \hline ATLAS & \ljets\ & $7~TeV$ & $4.7~fb^{-1}$ & $0.6 \pm 1.0$ &
$1.23 \pm 0.05$ \\ \cline{2-2} \cline{5-5} & dileptons & & & $5.7 \pm
2.4 (stat.) \pm 1.5 (syst.)$ & \\ \cline{1-2} \cline{4-5} & \ljets\ &
& $5~fb^{-1}$ & $0.4 \pm 1.0 (stat.) \pm 1.1 (syst.)$ & \\ \cline{2-2}
\cline{5-5} CMS & dileptons & & & $ 5.0 \pm 4.3 (stat.)
\ ^{+1.0}_{-3.9}\ (syst.)$ & \\ \cline{2-6} & \ljets\ & $8~TeV$ &
$19.7~fb^{-1}$ & $0.5 \pm 0.7 (stat.) \pm 0.6 (syst.)$ & $1.11 \pm
0.04$ \\ \hline
\end{tabular}
\caption{\label{LHC_results} \Ac\ measurements at the LHC
  \cite{ATLAS:ljets,ATLAS:dileptons,CMS:ljets,CMS:8tev}, unfolded to
  the parton level.}
\end{table}

\begin{table}
\renewcommand{\arraystretch}{1.3}
\begin{tabular}{l|c|c|c|c|c}
\hline \multicolumn{4}{c|}{Measurement} & Measured Value, \% &
\parbox{2.7cm}{Theoretical \\ Expectation\cite{Bernreuther:2012sx},\%}
\\ \hline ATLAS & dileptons & $7~TeV$ & $4.7~fb^{-1}$ & $2.3 \pm 1.2
(stat.) \pm 0.8 (syst.)$ & $0.55 \pm 0.02$\\ \cline{1-1} \cline{4-5}
CMS & & & $5~fb^{-1}$& $1.0 \pm 1.5 (stat.) \pm 0.6 (syst.)$ &
\\ \hline
\end{tabular}
\caption{\label{LHC_results_lepton} \Acll\ measurements at the LHC
  \cite{ATLAS:dileptons,CMS:dileptons}.}
\end{table}

\section{Tevatron Results}

\begin{figure}
\begin{minipage}[t]{.49\textwidth}
\includegraphics[width=\textwidth]{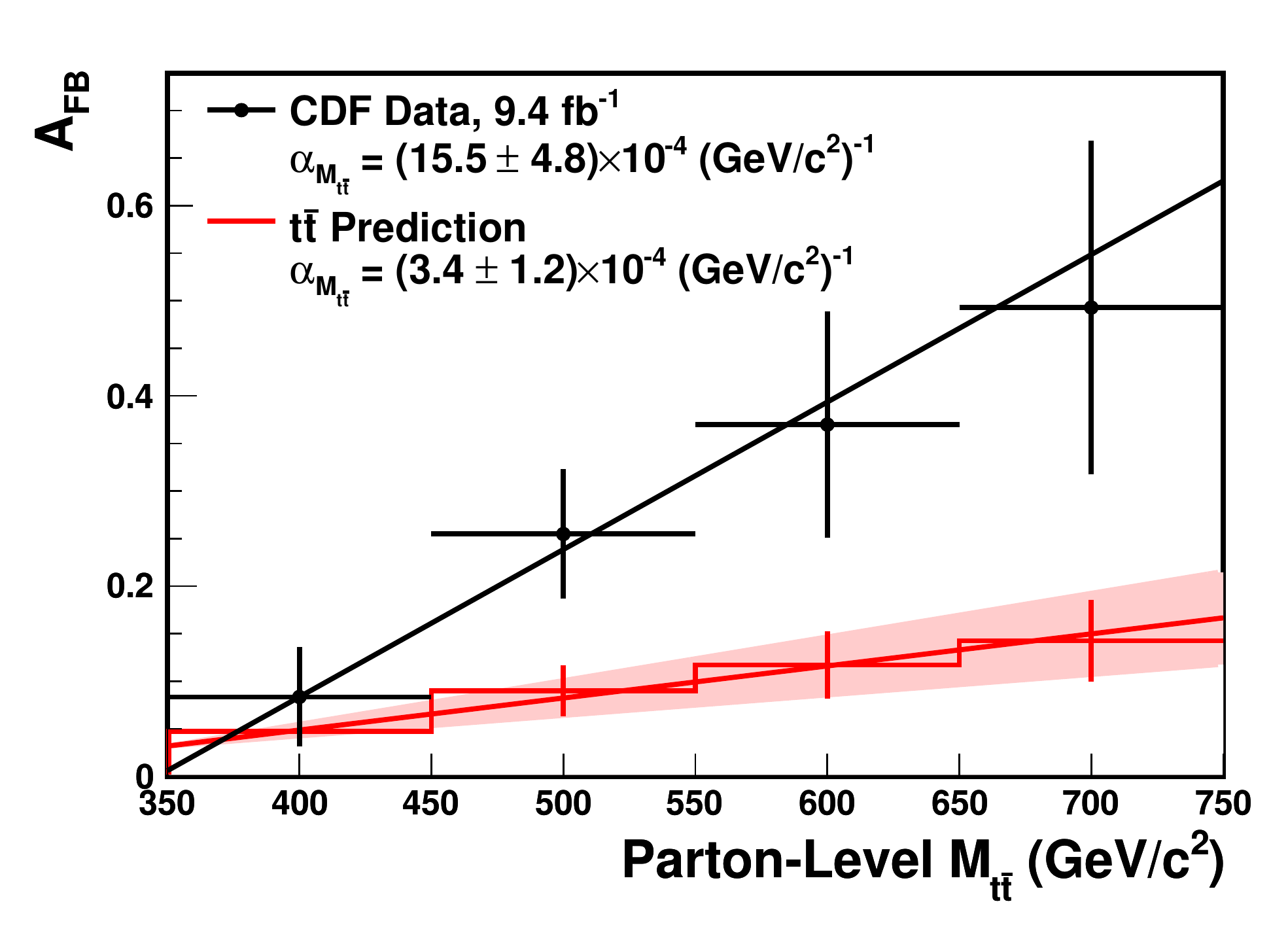}
\end{minipage}
\hfill
\begin{minipage}[t]{.49\textwidth}
\centerline{\includegraphics[width=\textwidth]{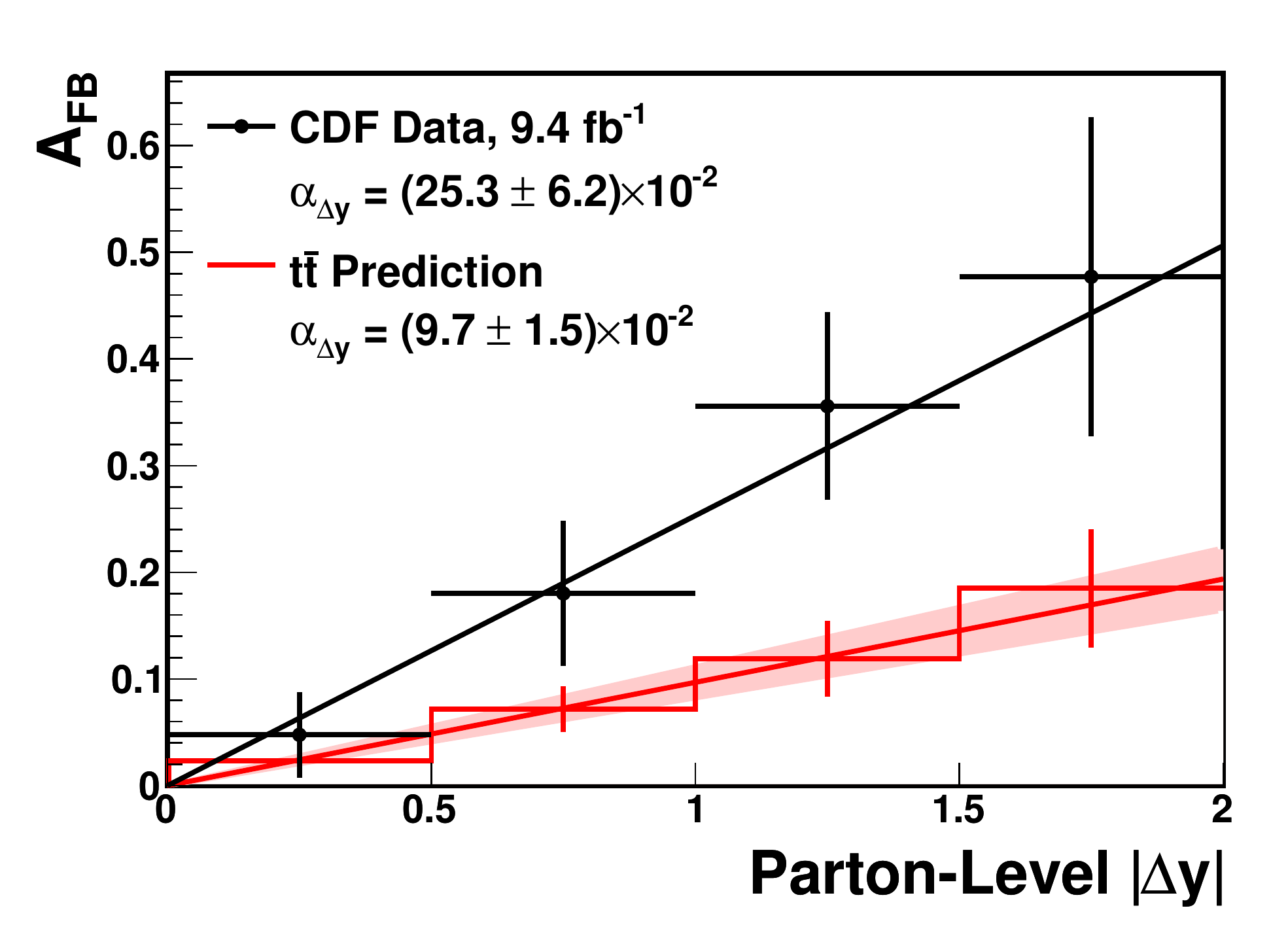}}
\end{minipage}
\caption{\label{Fig:cdf:ljets}\Afb\ asymmetry distribution as a
  function of $m_{t\bar{t}}$ and $|\Delta y|$ as measured by the CDF
  experiment in the \ljets\ final state~\cite{CDF:ljets:ttbar}.}
\end{figure}

For quite some times, measurements at the Tevatron were puzzling
because of the observed 2--3~standard deviations (SD) deviations
between the measured and expected \Afb\ asymmetries.
Table~\ref{Tevatron_ttbar_asymmetry} shows that the difference between
the most recent theoretical prediction and the current measurements of
the CDF and D0 experiments~\cite{CDF:ljets:ttbar,D0:ljets:ttbar} are
less than 2~SD.  In the same time, asymmetry measured at the high
$m_{t\bar{t}}$\ shows a moderate deviation from the expectation.  In
particular, CDF results on the measured slope of the \Afb\ asymmetry
as a function of $m_{t\bar{t}}$ (Fig.~\ref{Fig:cdf:ljets}) show a
2.4~SD deviation between measured slope and the expected one.  In the
same measurement $|\Delta y|$ dependence shows even large deviation at
the level of 2.8~SD, Fig.~\ref{Fig:cdf:ljets}.  The $|\Delta y|$
differential distribution of asymmetry has been also measured in a
different way in the CDF experiment.  The shape of the unfolded
$\cos(\theta)$ distribution has been fitted with the Legendre
polynomial series and it was found that the contribution to the
asymmetry of the first coefficient in series is different from the SM
expectation, see~\cite{CDF:ljets:differential} for the detailed
description.

The leptonic asymmetry \Al\ is measured by both experiments with a
full available statistics and demonstrates an agreement at the level of 
2~SD with the SM expectation (Table~\ref{Tevatron_al_asymmetry}), even if the CDF
measurement is slightly higher than the expectation.  It exists some
difficulty in the interpretation of the obtained results. It is
related to the fact, that leptonic asymmetries are measured in the
phase space limited by the acceptance $|y|$ cut and then extrapolated
to the full phase space.  These acceptance cuts are different in
different measurements, e.g. CDF \ljets\ measurement uses $|y|< 1.25$,
D0 \ljets\ measurement uses $|y|< 1.5$ and asymmetry in the D0
dilepton channel is measured within the acceptance cut of $|y|< 2.0$.
The extrapolation procedure is model dependent and done in a different
way in a different measurements.  Currently both experiments are
working on the combination of measurements and defining the most
appropriate extrapolation procedure.

\begin{figure}
\centerline{\includegraphics[width=.5\textwidth]{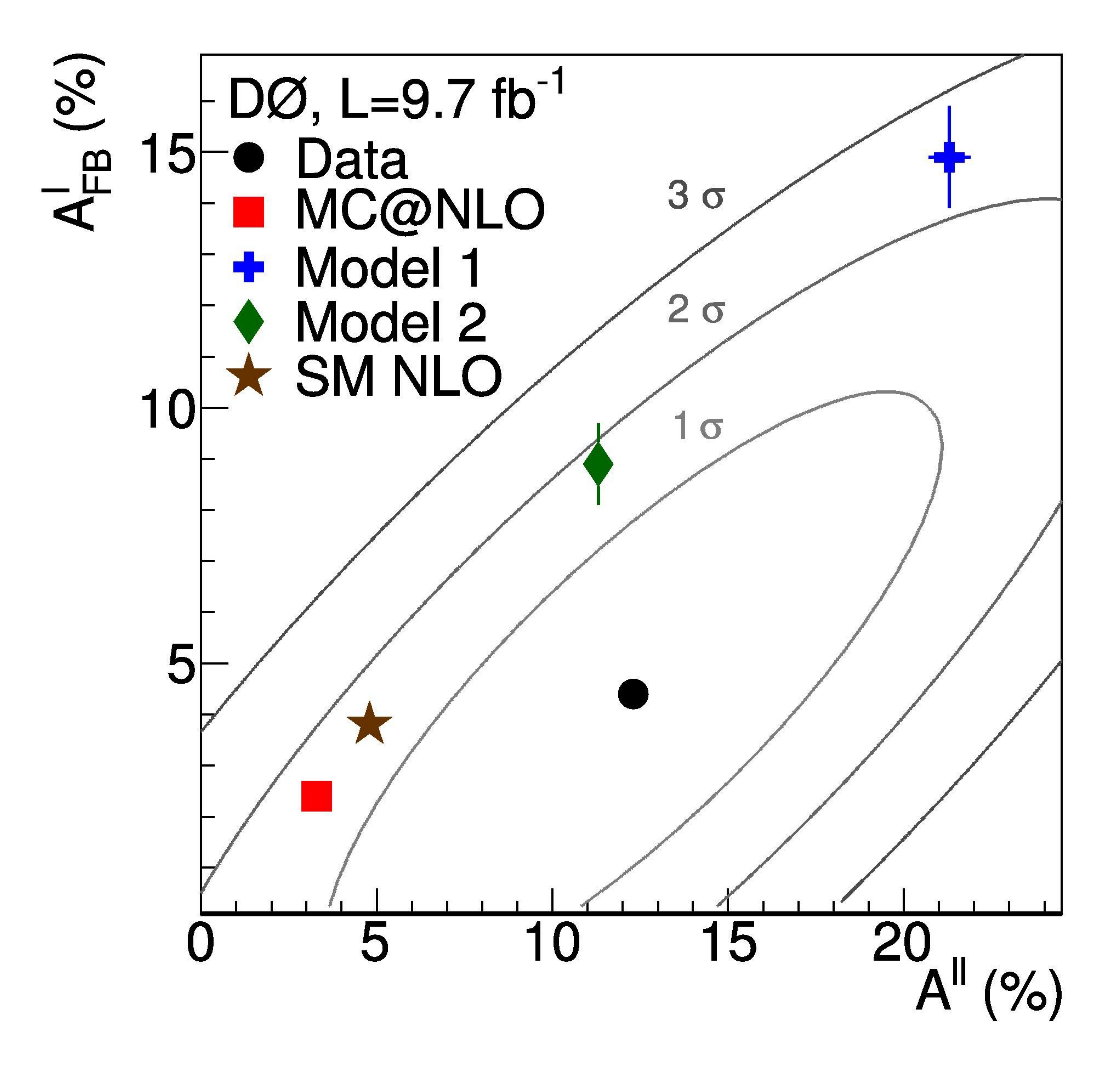}}
\caption{\label{Fig:D0:dileptons}\Al\ vs \All\ asymmetry as measured
  by the D0 experiment in the dilepton final
  state~\cite{D0:dileptons,Chapelain}.}
\end{figure}

The dilepton final state gives an unique possibility to make a
measurement of \All\ asymmetry. The D0 analysis measured it to be
equal $12.3 \pm 5.4 (stat.) \pm 1.5 (syst.)$~\cite{D0:dileptons,Chapelain} which is 
in agreement with the theoretical prediction $4.8\pm 0.4$~\cite{Bernreuther:2012sx}. In
addition, in this analysis the correlation between \All\ and
\Al\ measurements has been studied, see Fig~\ref{Fig:D0:dileptons},
and the ratio of these two asymmetries has been found to be $R =
A_\ell / A_{FB}^{\ell\ell} = 0.36\pm0.20$ which is 2~SD away from the
expectation, which could be estimated using the predicted values of
\All\ and \Al\ in~\cite{Bernreuther:2012sx}: $R_{th} = 3.8 / 4.8 \sim
0.8$. For further discussion about this measurement
see~\cite{Chapelain}.

\begin{table}
\renewcommand{\arraystretch}{1.3}
\begin{center}
\begin{tabular}{l|c|c|c}
\hline \multicolumn{2}{c|}{Measurement} & Measured Value, \% &
\parbox{2.9cm}{Theoretical \\ Expectation \cite{Bernreuther:2012sx},
  \%} \\ \hline CDF & $9.4~fb^{-1}$ & $16.4\pm4.7$ & $8.8 \pm 0.6$
\\ \cline{1-4} D0 & $5.4~fb^{-1}$ & $19.6\pm6.5$ & \\ \hline
\end{tabular}
\caption{\label{Tevatron_ttbar_asymmetry} \Afb\ measurements in the
  \ljets\ final state at the Tevatron
  \cite{CDF:ljets:ttbar,D0:ljets:ttbar}, unfolded to the parton
  level.}
\end{center}
\end{table}
\begin{table}
\renewcommand{\arraystretch}{1.3}
\begin{center}
\begin{tabular}{l|c|c|c|c}
\hline \multicolumn{3}{c|}{Measurement} & Measured Value, \% &
\parbox{2.9cm}{Theoretical \\ Expectation \cite{Bernreuther:2012sx},
  \%} \\ \hline CDF & \ljets\ & $9.4~fb^{-1}$ & $9.4^{+3.2}_{-2.9}$ &
$3.8 \pm 0.6$ \\ \cline{1-4} D0 & \ljets\ ($|\eta| < 1.5$) &
$9.7~fb^{-1}$ & $4.7 \pm 2.3 (stat.)\ ^{+1.1}_{-1.4}\ (syst.)$ &
\\ \cline{1-4} D0 & dileptons & $9.7~fb^{-1}$ & $4.4 \pm 3.7 (stat.)
\pm 1.1 (syst.) $ & \\ \hline
\end{tabular}
\caption{\label{Tevatron_al_asymmetry}\Al\ measurements at the
  Tevatron \cite{CDF:ljets:leptonic,D0:ljets:leptonic, D0:dileptons}.
  CDF \ljets\ and D0 dileptons measurements are extrapolated to the
  full phase space, but D0 \ljets\ measurement is limited to the
  acceptance $|\eta| < 1.5$.}
\end{center}
\end{table}


\clearpage

\section{Conclusion}

During last several years Tevatron measurements of the asymmetry show
an intriguing deviation from the SM calculation.  The current
measurements of the inclusive \ttbar\ asymmetry from the CDF and from the D0 experiments
don't show any strong deviation from the recent SM calculations, but the
asymmetry measurement as a function of $m_{t\bar{t}}$ or $|\Delta y|$
show a significant deviation from the SM at the level more than 2
standard deviations. In the same time D0 didn't yet analyzed the full
available statistics and hence the final conclusion about the
\ttbar\ asymmetry from the Tevatron is still to come.  The leptonic
asymmetry measurements at the Tevatron deviate less than 2~SD from the 
 SM model predictions, but results are still need to be combined to have a more precise
conclusion about the level of agreement with the expectation.

\begin{figure}
\begin{minipage}[t]{0.49\textwidth}
\includegraphics[width=\textwidth]{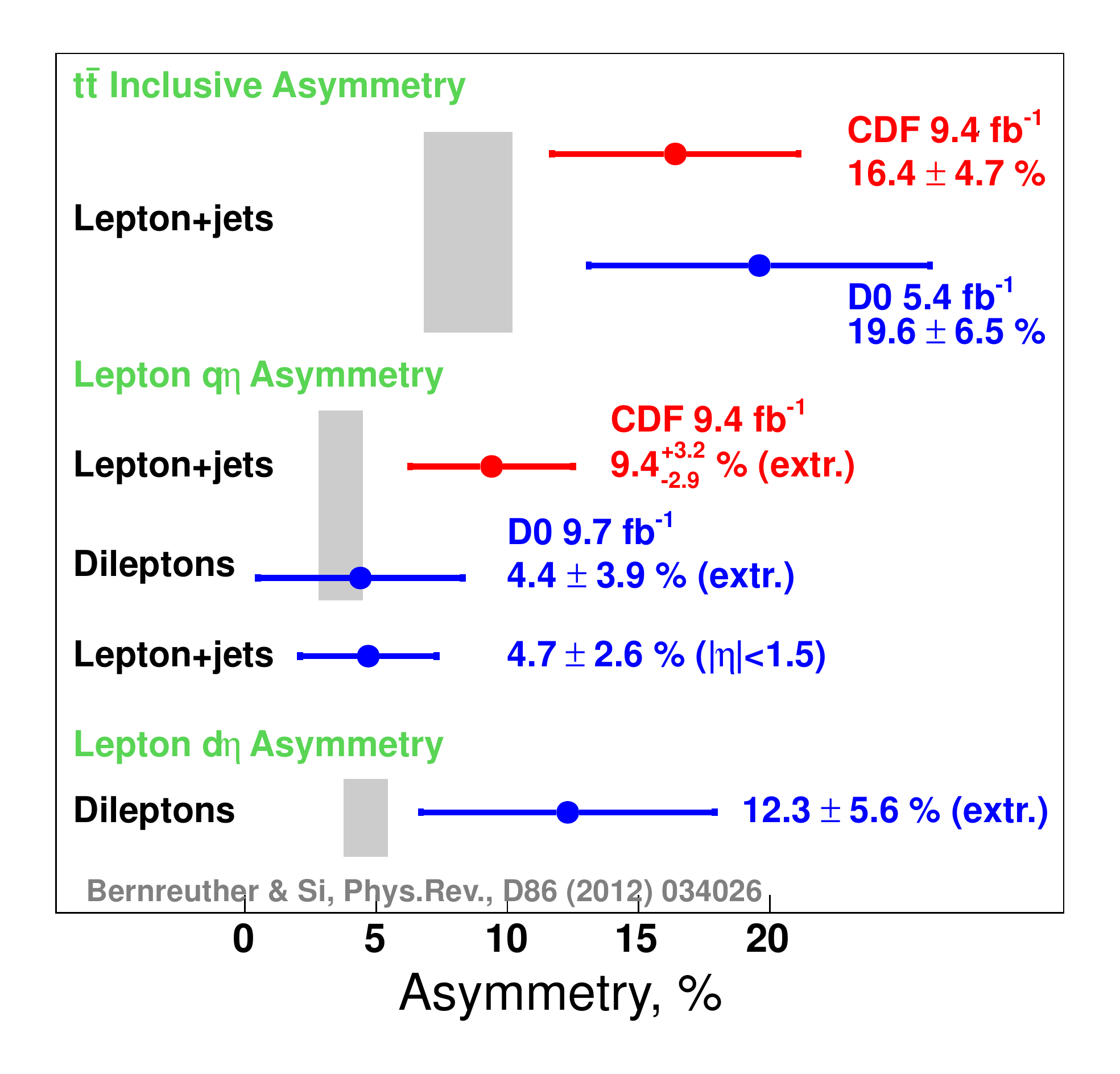}
\end{minipage}
\hfill
\begin{minipage}[t]{0.49\textwidth}
\centerline{\includegraphics[width=\textwidth]{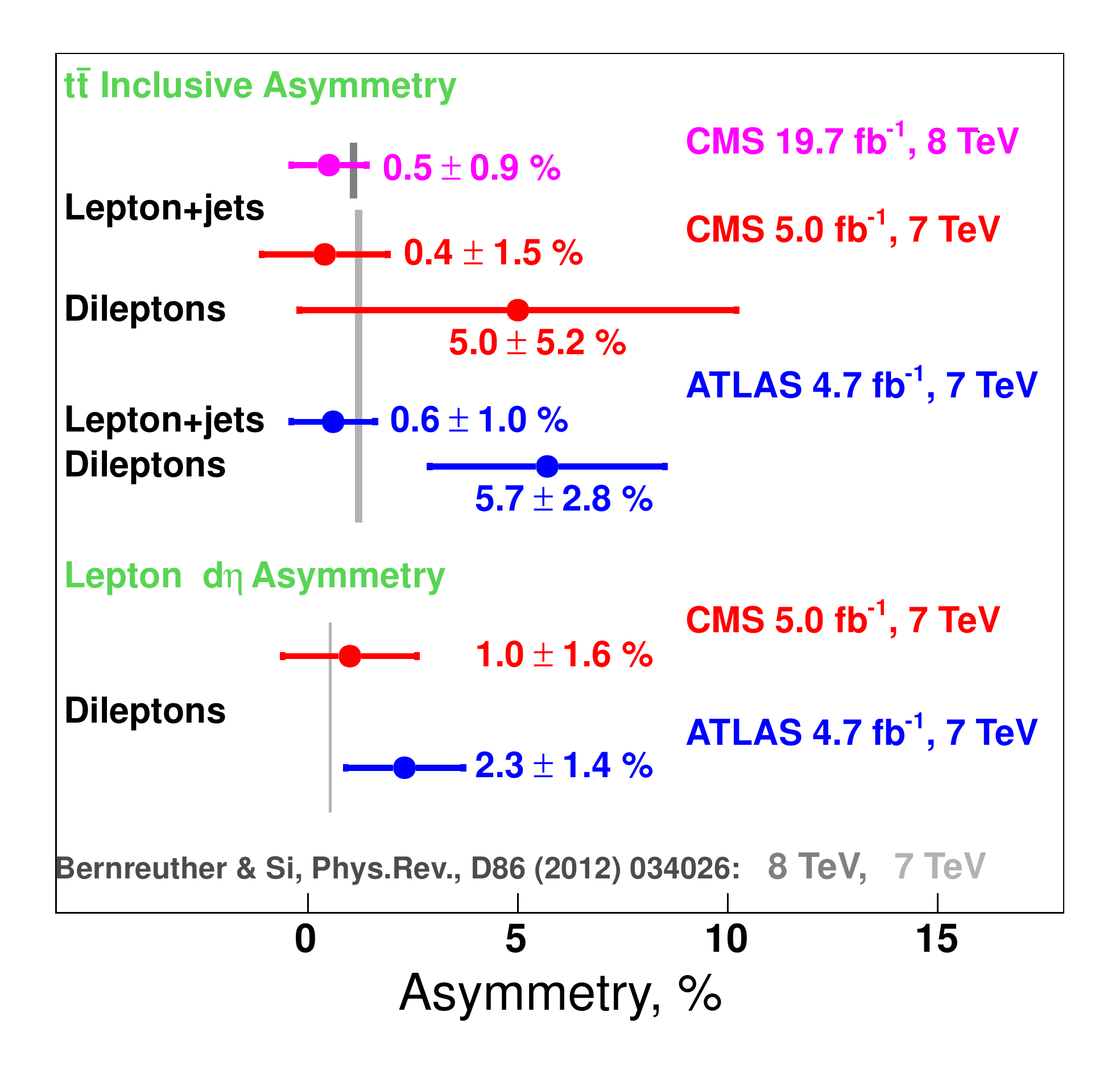}}
\end{minipage}
\caption{\label{Fig:measurements} Summary of the asymmetry
  measurements at the Tevatron and LHC.}
\end{figure}

Measurements at the LHC don't show any deviation from the SM
prediction, but the expected asymmetry is very low and the current
precision of the measurements is about 1\%, comparable with the
expected asymmetry.  At the LHC the most interesting direction of
study is a measurement of the differential asymmetry as a function of
the velocity or invariant \ttbar\ mass.  The large statistics
accumulated at the LHC make possible the precise measurements in
regions of the phase space where both SM and non-SM asymmetries are
expected to be large. For the moment, no deviation from the
expectations were found.  Fig.~\ref{Fig:measurements} summaries
current measurements of the inclusive asymmetries both at the Tevatron
and LHC and compares them with the expected SM values.


\begin{footnotesize}

\end{footnotesize}



\begin{thebibliography}{99}

\bibitem{Bernreuther:2012sx} W.~Bernreuther and Z.~-G.~Si, 
  quark and leptonic charge asymmetries for the Tevatron and LHC,''
  Phys.\ Rev.\ D {\bf 86} (2012) 034026 [arXiv:1205.6580 [hep-ph]].

\bibitem{Westhoff:2013ixa} S.~Westhoff, in this proceedings, ``Top
  Charge Asymmetry -- Theory Status Fall 2013,'' arXiv:1311.1127
  [hep-ph].  


\bibitem{Falkowski:2012cu} A.~Falkowski, M.~L.~Mangano, A.~Martin,
  G.~Perez and J.~Winter, 
  forward--backward asymmetry with a lepton-based handle,''
  Phys.\ Rev.\ D {\bf 87} (2013) 034039 [arXiv:1212.4003 [hep-ph]].
  of 14 Nov 2013

\bibitem{ATLAS:ljets} The ATLAS collaboration, 
  top quark pair production charge asymmetry in proton-proton
  collisions at $sqrt{s}=7$ TeV using the ATLAS detector,''
  ATLAS-CONF-2013-078.  

\bibitem{ATLAS:dileptons} The ATLAS collaboration,
  ATLAS-CONF-2012-057.

\bibitem{CMS:ljets} S.~Chatrchyan {\it et al.}  [CMS Collaboration],
  asymmetry in proton-proton collisions at 7 TeV,'' Phys.\ Lett.\ B
  {\bf 717} (2012) 129 [arXiv:1207.0065 [hep-ex]].  
  ARXIV:1207.0065;
  2013

\bibitem{CMS:dileptons} The CMS Collaboration, 
  measurement in dileptons at 7 TeV,'' CMS-PAS-TOP-12-010.

\bibitem{CMS:8tev} CMS Collaboration [CMS Collaboration],
  at 8 TeV,'' CMS-PAS-TOP-12-033.  

\bibitem{AguilarSaavedra:2012va} J.~A.~Aguilar-Saavedra and A.~Juste,
  Phys.\ Rev.\ Lett.\ {\bf 109} (2012) 211804 [arXiv:1205.1898
    [hep-ph]].\\ J.~A.~Aguilar-Saavedra, A.~Juste and F.~Rubbo,
  (2012) 92 [arXiv:1109.3710 [hep-ph]].

\bibitem{CDF:ljets:ttbar} T.~Aaltonen {\it et al.}  [CDF
  Collaboration], 
  production asymmetry and its dependence on event kinematic
  properties,'' Phys.\ Rev.\ D {\bf 87} (2013) 092002 [arXiv:1211.1003
    [hep-ex]].

\bibitem{D0:ljets:ttbar} V.~M.~Abazov {\it et al.}  [D0
  Collaboration], 
  production,'' Phys.\ Rev.\ D {\bf 84} (2011) 112005 [arXiv:1107.4995
    [hep-ex]].

\bibitem{CDF:ljets:differential} T.~Aaltonen {\it et al.}  [CDF
  Collaboration], 
  Phys.\ Rev.\ Lett.\ {\bf 111}
  (2013) 182002 [arXiv:1306.2357 [hep-ex]].

\bibitem{CDF:ljets:leptonic} T.~A.~Aaltonen {\it et al.}  [CDF
  Collaboration], 
  events produced in ppbar collisions at sqrt(s)=1.96 TeV,''
  Phys.\ Rev.\ D {\bf 88} (2013) 072003 [arXiv:1308.1120 [hep-ex]].


\bibitem{D0:ljets:leptonic} V.~M.~Abazov {\it et al.}  [D0
  Collaboration], Conference note 6381, July 2013

\bibitem{D0:dileptons} V.~M.~Abazov {\it et al.}  [D0 Collaboration],
  produced in dilepton ttbar final states in $p\bar p$ collisions at
  $\sqrt{s}=1.96$ TeV,'' Accepted by Phys. Rev. D (2013),
  [arXiv:1308.6690 [hep-ex]].



\bibitem{Chapelain} A.~Chapelain for the D0 Collaboration, in this
  proceedings, ``Measurement of the leptonic ttbar charge asymmetry in
  the dilepton channel with the D0 detector,'' arXiv:1311.6731
  [hep-ex].

\end{thebibliography}
\end{document}